\documentclass[journal,nofonttune,twoside]{IEEEtran}

\usepackage[separate-uncertainty = true, multi-part-units=single]{siunitx}
\DeclareSIUnit\pe{p.e.}
\usepackage{subfigure}
\usepackage{booktabs}
\usepackage{color}
\usepackage{amsmath}
\usepackage{breqn}
\usepackage[mathlines,switch]{lineno}
\usepackage[]{footmisc}
\usepackage[breaklinks]{hyperref}
\usepackage{soul}

\usepackage{authblk}

\newcolumntype{L}[1]{>{\raggedright\let\newline\\\arraybackslash\hspace{0pt}}m{#1}}
\newcolumntype{C}[1]{>{\centering\let\newline\\\arraybackslash\hspace{0pt}}m{#1}}
\newcolumntype{R}[1]{>{\raggedleft\let\newline\\\arraybackslash\hspace{0pt}}m{#1}}

\ifCLASSINFOpdf
\usepackage[pdftex]{graphicx}
\DeclareGraphicsExtensions{.pdf,.jpeg,.png}
\else

\fi

\usepackage{url}
\IEEEoverridecommandlockouts

\begin{document}
\title{VUV-sensitive Silicon Photomultipliers for Xenon Scintillation Light Detection in nEXO}

\author{A.~Jamil, T.~Ziegler, P.~Hufschmidt, G.~Li, L.~Lupin-Jimenez, T.~Michel, I.~Ostrovskiy, F.~Reti\`ere, J.~Schneider, M.~Wagenpfeil
% ...
\thanks{Manuscript received June 8, 2018; revised August 29, 2018. Please see Acknowledgment section for author affiliations.}}

\markboth{Prepared for submission to IEEE TRANSACTIONS ON NUCLEAR SCIENCE}{Jamil \MakeLowercase{\textit{et al.}}: VUV-sensitive Silicon Photomultipliers for Xenon Scintillation Light Detection in nEXO}

\maketitle

\begin{abstract}
Future tonne-scale liquefied noble gas detectors depend on efficient light detection in the VUV range. In the past years Silicon Photomultipliers (SiPMs) have emerged as a valid alternative to standard photomultiplier tubes or large area avalanche photodiodes. The next generation double beta decay experiment, nEXO, with a 5 tonne liquid xenon time projection chamber, will use SiPMs for detecting the {\SI{175}{\nano\meter}} xenon scintillation light, in order to achieve an energy resolution of $\boldsymbol{\sigma}/\boldsymbol{Q_{\beta\beta}} = \SI{1}{\percent}$. This paper presents recent measurements of the VUV-HD generation SiPMs from Fondazione Bruno Kessler in two complementary setups. It includes measurements of the photon detection efficiency with gaseous xenon scintillation light in a vacuum setup and dark measurements in a dry nitrogen gas setup. We report improved photon detection efficiency at \SI{175}{\nano\metre} compared to previous generation devices, that would meet the criteria of nEXO. Furthermore, we present the projected nEXO detector light collection and energy resolution that could be achieved by using these SiPMs.
\end{abstract}


\IEEEpeerreviewmaketitle

\section{Neutrino-less double beta decay and nEXO}

\IEEEPARstart{N}{}eutrino-less double beta decay ($0\nu\beta\beta$) is a hypothetical nuclear decay where two neutrons decay into two protons and two electrons are emitted but no anti-neutrinos are present in the final state. The observation of this process would have a fundamental impact on the Standard Model of Particle Physics, specifically showing a violation of lepton number conservation ($|\Delta L|\;=2$), and would imply that the neutrino is a Majorana fermion~\cite{Majorana}, independently of the actual process enabling the decay~\cite{schechter-valle}. Furthermore, the half-life of the decay would shed light on the absolute neutrino mass scale~\cite{gomez}.
\parskip0pt

\begin{figure}[t]
\centering
\includegraphics[width=0.9\columnwidth]{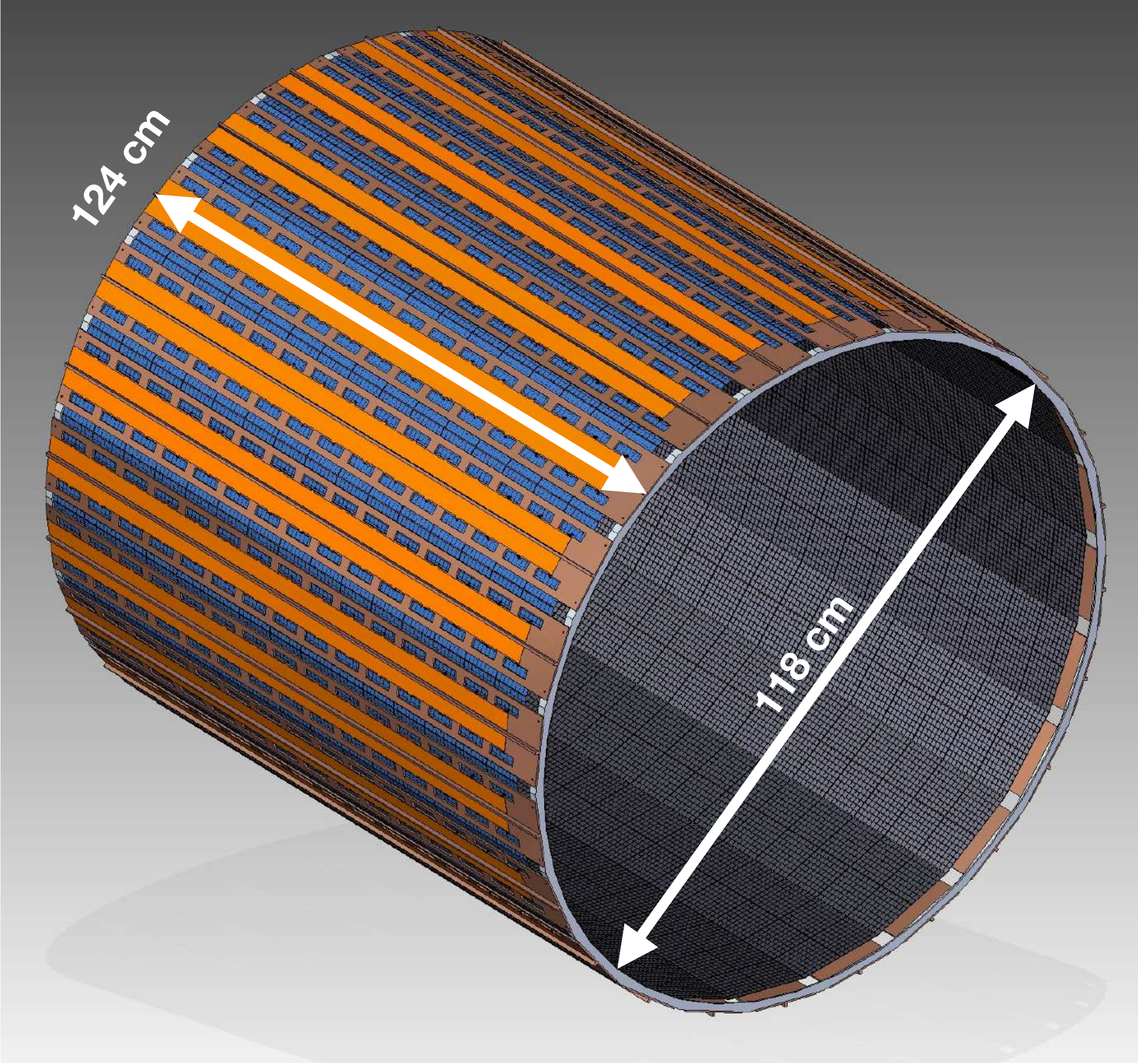}
\caption{Conceptual arrangement of SiPMs inside the nEXO TPC. The SiPMs will be grouped into tiles of $\num{8}\times\num{8}$ dies with \num{30} tiles mounted onto staves, and a total of \num{24} staves to cover the whole lateral surface. The full assembly is \SI{124}{\centi\meter} in height and \SI{118}{\centi\meter} in diameter and will incorporate about \SI{4}{\meter\squared} of SiPMs.}
\label{fig:nexo}
\end{figure}

The nEXO collaboration plans to build a cylindrical single-phase time projection chamber (TPC) filled with \num{5} tonnes of liquid xenon (LXe), with \SI{90}{\percent} enrichment in $^{136}$Xe~\cite{sensitivity}. nEXO takes advantage of the experience from its predecessor \mbox{EXO-200}~\cite{exo}, but will incorporate new light and charge detectors~\cite{tiles}. Together with cold electronics sitting inside the LXe, this allows nEXO to achieve an energy resolution of $\sigma/Q_{\beta\beta} = \SI{1}{\percent}$ for the $0\nu\beta\beta$ decay of $^{136}\mathrm{Xe}$ (\SI{2458.07\pm 0.31}{\kilo\electronvolt}~\cite{qvalue,qvalue2}). In particular, instead of the \mbox{EXO-200} Large Area Avalanche Photo-diodes (LAAPDs), nEXO will use Silicon Photomultipliers (SiPMs) for the detection of xenon scintillation light. The SiPMs will fully cover the lateral surface of the cylinder with a total photo-sensitive area of about \SI{4}{\meter\squared}, as shown in Figure~\ref{fig:nexo}. The devices will be immersed in LXe and placed in the high field region behind the field shaping rings of the TPC field cage~\cite{tamer}. The performance of SiPMs has improved significantly over the past decade and they are especially interesting because of their high gain, on the order of \num{e6}, and their single photon resolution capability.

The half-life sensitivity of nEXO to the $0\nu\beta\beta$ decay of $^{136}\mathrm{Xe}$ is projected to be \SI{9.5e27}{yr} for \SI{90}{\percent} C.L.\ after \num{10} years of data taking~\cite{sensitivity}. According to recent work~\cite{Agostini}, nEXO is one of the planned double beta decay experiments with the best discovery potential.
In order to achieve the anticipated energy resolution and sensitivity, the nEXO collaboration has started an extensive characterization campaign of SiPMs to find the optimal candidate~\cite{Ostrovskiy} and is working together with different vendors, in particular Fondazione Bruno Kessler (FBK)~\cite{fbk} and Hamamatsu Photonics K.K.~\cite{hamamatsu}. The most important parameters to consider are:
\begin{itemize}
\item The Photon-detection efficiency (PDE). This is defined as {
\begin{linenomath*}
\begin{eqnarray}
\mathrm{PDE_{\rm measured}} & = &\mathrm{PDE}\cdot (1-r(\theta)) \nonumber\\
& = & \epsilon_{\rm geo} \cdot P_{\rm trig} \cdot \mathrm{QE} \cdot (1-r(\theta))
\end{eqnarray}
\end{linenomath*}}
where $P_{\rm trig}$ is the probability that a photoelectron triggers an avalanche, QE is the quantum efficiency and $\epsilon_{\rm geo}$ is the ratio of photo-sensitive area to overall surface area. In general, the PDE also depends on the angle-dependent reflectance $r(\theta)$ of the SiPM surface. Important for the energy resolution is the overall light collection efficiency
\begin{linenomath*}
\begin{equation}
\epsilon_0 = \mathrm{PDE} \cdot \mathrm{PTE}
\end{equation}
\end{linenomath*}
where PTE is the photon transport efficiency, which is defined as the fraction of photons that are absorbed in the SiPMs relative to the number of originally emitted photons. Both a high PDE and high reflectivity of components within the TPC (high PTE) are crucial to minimize the loss of photons. More than half of the light impinging on the SiPMs is expected to be reflected due to the mismatch in refractive indexes between the SiPM surface and LXe. A detailed understanding of reflectance as a function of the angle of incidence on the SiPM surface is necessary to increase the quality of the simulations at reproducing $\epsilon_0$. According to current simulations, a PDE of at least \SI{15}{\percent} and a PTE of \SI{20}{\percent} is sufficient to achieve a $\sigma/Q_{\beta\beta} = \SI{1}{\percent}$ energy resolution.

\item Correlated avalanches. These are secondary avalanches that are not triggered directly by an initially absorbed photon or a {dark event}. We can define two sub-categories (that themselves can be further sub-categorized): optical crosstalk and afterpulsing. The former process is due to photon emission by electrons in collisions during the avalanche that could trigger another avalanche in another microcell. The latter effect occurs when an electron in the avalanche region is trapped in a lattice defect and triggers an additional avalanche after being released. The two processes can be distinguished by studying the time distribution of events {relative to the prompt pulse (triggered pulse)} (see Figure~\ref{fig:pe_distribution} in section~\ref{subsec:sipm_dark}). In order to reach nEXO design performance, the fraction of correlated avalanches of both types combined per parent avalanche within a time window of \SI{1}{\micro\second} should be at most \SI{20}{\percent}. The higher this value, the larger the contribution to the overall fluctuations of detected photons will be, resulting in worse energy resolution.

\item Gain and capacitance per unit area. The signal-to-noise ratio of the front-end electronics for single-photoelectrons depends on both SiPM gain and capacitance per unit area~\cite{fabris_talk,fabris_thesis}. This, under a power consumption constraint, sets an upper limit on the total area one channel of the front-end electronics can read out. Since there is little information on the geometrical origin of the scintillation light, complexity is minimized by grouping several SiPMs into a single readout channel and, at present, it is assumed that \SI{6}{\centi\meter\squared} will be read out by one channel. This can be achieved as long as the specific capacitance of the SiPMs is below \SI{50}{\pico\farad\per\milli\meter\squared} and the sum of gain fluctuations and electronics noise is smaller than \SI{0.1}{\pe} {In addition, prior to the assembly SiPMs with similar behavior in terms of gain and breakdown voltage will be identified and grouped together.} 
\end{itemize}

This work focuses on the VUV-HD SiPMs from FBK with enhanced PDE at {\SI{175}{\nano\meter}}. These SiPMs have a so-called low-field (LF) and standard-field (STD) version, which differ in their doping profile. Both have a surface area of $\num{5.56}\times\SI{5.96}{\milli\metre\squared}$ and are made up of an array of \num{4} smaller SiPMs, which have been connected in parallel. {These devices have a microcell pitch of \SI{30}{\micro\meter} and total of $\sim$700 microcells per \SI{1}{\milli\meter\squared}. The breakdown voltage of the STD and LF devices were measured to be \SI{22.8}{\volt} and \SI{28.8}{\volt} at \SI{169}{\kelvin}, respectively.} We present results from complementary setups at Stanford and Erlangen. The former one is used for efficiency measurements that help identifying potential SiPMs candidates. Measurements with the latter setup are aimed for better understanding of the devices properties in order to improve future SiPMs together with vendors.
These results are then used to infer the performance of the nEXO detector in terms of the achievable energy resolution.

\section{Hardware and Signal Processing}
\subsection{Stanford Setup}

\begin{figure}[t]
\centering
\includegraphics[width=.9\columnwidth]{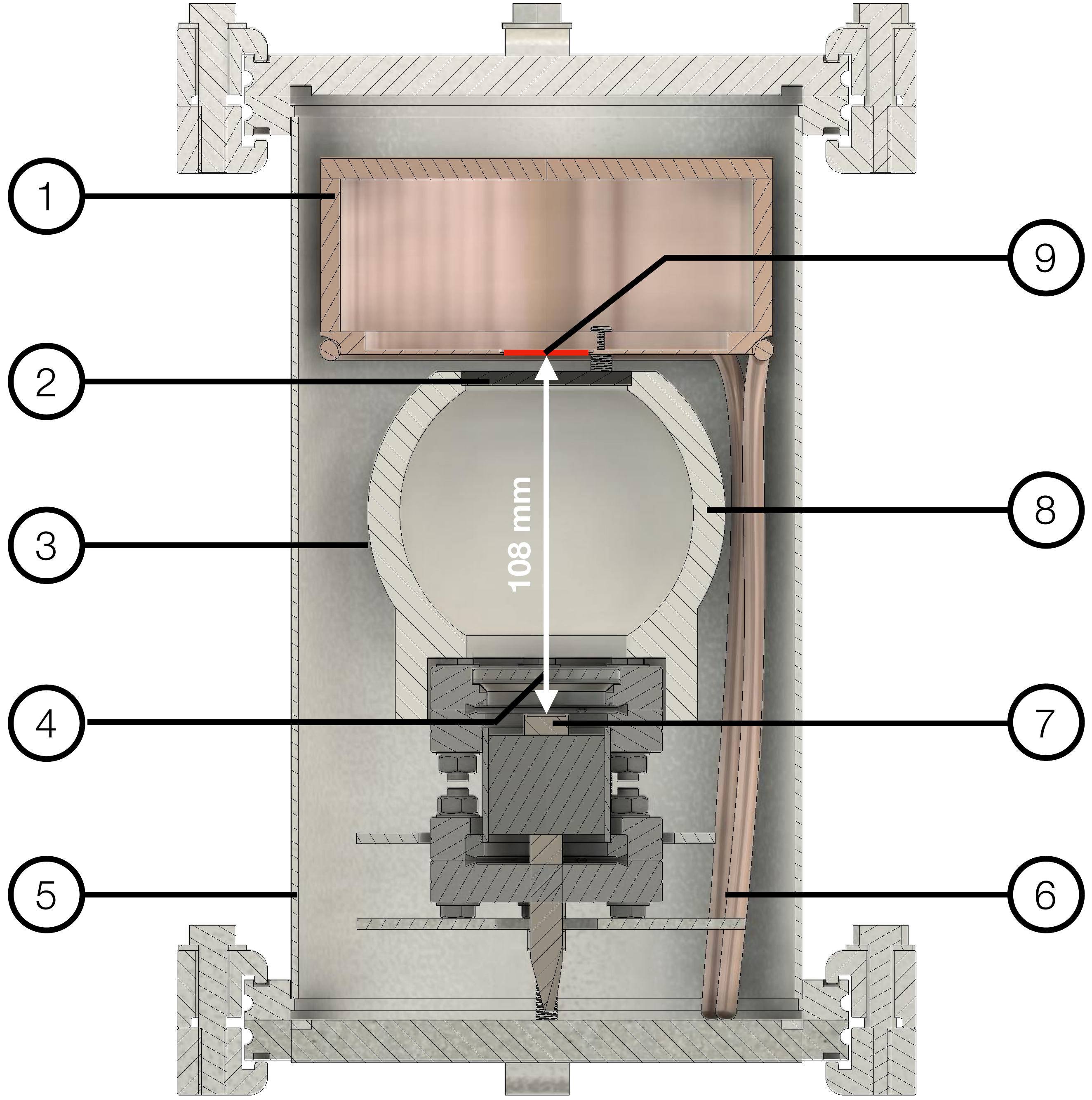}
\caption{Cross-sectional view of the SiPM test setup at Stanford. The custom made scintillation light source (7) is at the bottom of the vacuum chamber (5). The detector cage (1) and the detectors (9) are cooled through a copper tube (6), that is brazed onto it and flushed with liquid nitrogen boil-off gas. The scintillation light that leaves the optical window (4) is confined within a PTFE sphere (8) to eliminate parasitic reflections off materials inside the vacuum chamber. In addition, a wavelength bandpass filter (2) is placed on top of the PTFE sphere. For a calibration of the detectors an optical fiber (3), that is connected to an LED outside of the vacuum chamber, is mounted to outside of the PTFE sphere (not shown in figure).}
\label{fig:stanford_setup}
\end{figure}
The Stanford group operates a test setup for characterizing SiPMs in vacuum with Xe scintillation light. It was optimized for efficiently determining the PDE of \SI{\sim 1}{\centi\metre\squared} size SiPMs. Figure~\ref{fig:stanford_setup} shows a cross-sectional view of the vacuum chamber (5), containing the cooling system, the source and the detector cage. The light source is custom made and consists of a quartz-stainless steel capsule filled with a xenon atmosphere at about \SI{1}{bar}. Inside the capsule, $^{252}$Cf, electroplated onto a platinum surface (7), produces xenon scintillation light after an $\alpha$-decay or spontaneous fission. The photons leave the source assembly through a quartz window (4). {A cavity made out of PTFE (known to be a good reflector in the VUV range) is interposed between the Xe scintillation source and the photodetectors, improving the light collection, while at the same time avoiding possible inconsistencies and systematics due to light scattering by the many complex materials in the vacuum chamber. The spherical shape of the reflector does not have a specific function in the current version of the setup}. In addition, an optical bandpass filter (2) with a transmission curve centered at \SI{180}{\nano\metre} and a width of \SI{40}{\nano\metre} FWHM was placed at the top opening on the detector side of the PTFE sphere. The bandpass filter eliminates photons other than the {\SI{175}{\nano\meter}} xenon scintillation light -- e.g.\ PTFE's re-emission (conversion, fluorescence) at longer wavelengths and the sub-dominant infra-red component of the xenon scintillation spectrum. An optical fiber (3) allows the calibration of the reference PMT and the SiPMs (9) with a light source of varying intensity and wavelength. {The fiber is mounted to the outside of the PTFE sphere (not shown in figure) and connected to an LED located outside the vacuum chamber}. Throughout this work we have used a blue LED for the calibration.

The detectors are placed inside a copper box (1) at the top of the vacuum chamber with a few mm distance to the optical bandpass filter. Since the setup only allows measurement of one detector at a time, different insets were made to allow positioning the reference PMT and the different SiPMs at an equal distance to the light source. In order to avoid electronics pick-up, PTFE spacers were used to electrically isolate the copper box from the rest of the vacuum chamber. Copper pipes (6) are brazed to the bottom perimeter of copper box, providing cooling by a flow of nitrogen boil-off gas. The temperature is controlled by resistive heaters, attached to the ingoing pipe and controlled by Omega PID controllers~\cite{omega}, to allow stable measurements at the LXe temperature in nEXO at around \SI{169}{\kelvin}, with a variation of \SI{0.5}{\kelvin}. The vacuum achieved by this setup is better than \SI{1E-6}{\milli\bar}. The system is designed for fast cooldown from room temperature and can reach stable operations in \SI{\sim 1.5}{\hour}.

\subsection{Erlangen Setup}

\begin{figure}[t]
\centering
\includegraphics[width=.8\columnwidth]{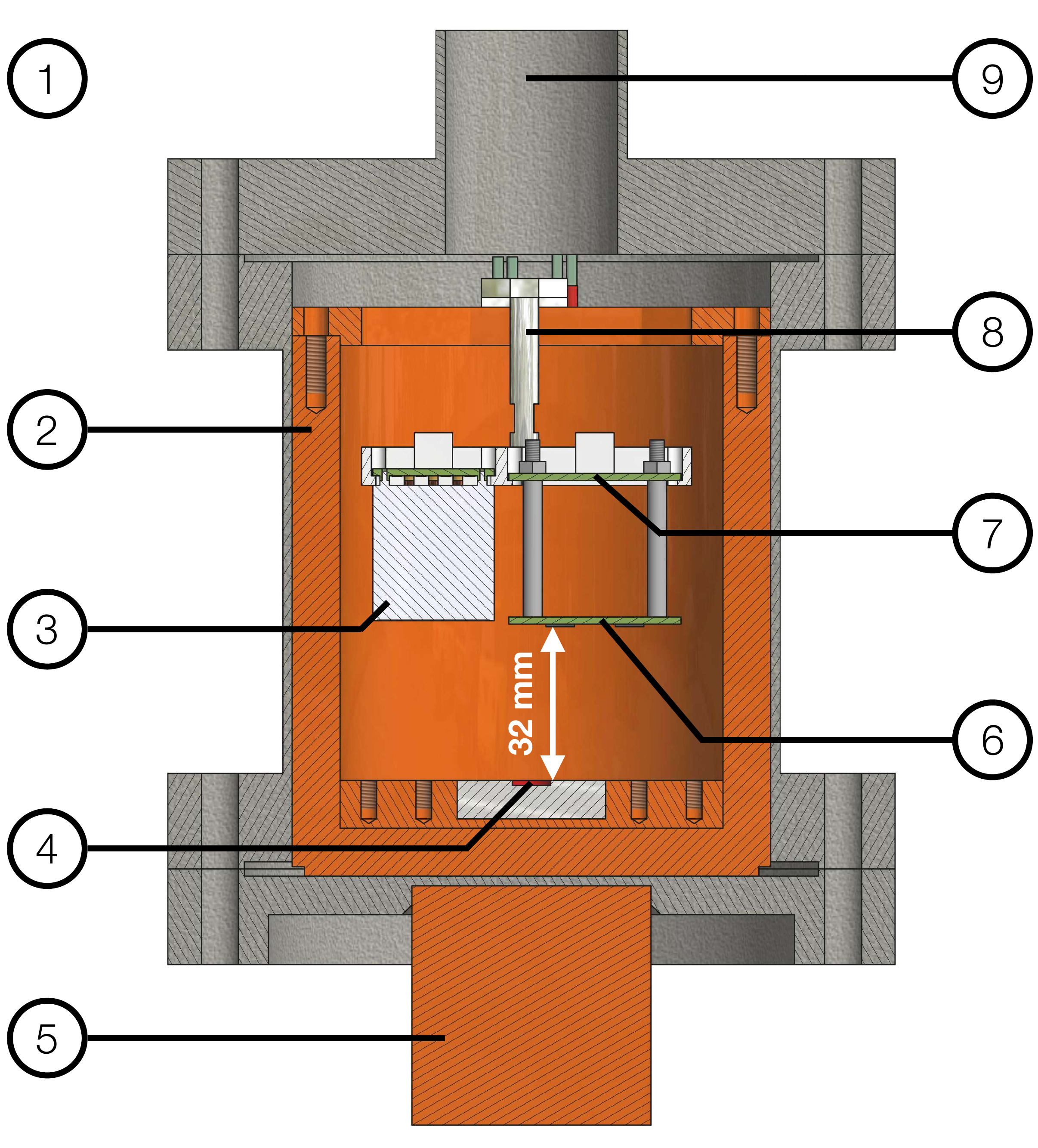}
\caption{Cross-section of the Erlangen SiPM test setup showing the inner vacuum containment (1). It comprises the copper cup (2) containing the xenon, the cooling finger (5), both PMT (3) and SiPM stack, combining SiPM board (6) and amplification board (7), the PTFE detector holder (8) and the tube connector to the electrical feedthroughs on top (9). The $\alpha$-source (4) can be placed inside an inset at the bottom of the copper cup.}
\label{fig:erlangen_setup}
\end{figure}

The Erlangen SiPM test setup consists of an inner xenon cell contained in an outer vacuum vessel (1). The xenon cell incorporates the detectors used for dark and xenon scintillation light measurements. Figure~\ref{fig:erlangen_setup} shows a cross-sectional view where one can see the SiPM stack of two printed circuit boards on the right and a VUV-sensitive PMT (3), R8520-406 from Hamamatsu, on the left. The SiPM stack consists of the SiPM carrier board (6) and a custom made preamp board (7). Both PMT and SiPM stacks are held in place by a PTFE structure (8) which hangs down from a copper bracket at the top of the xenon cell, thus having the detectors facing downwards. The whole xenon cell is enclosed by a copper cup (2) with a wall thickness of \SI{1}{\centi\metre} for thermal conductivity reasons. Besides the detector assembly, the xenon cell also contains a copper inlet at the bottom in which an $\alpha$-source (4) can be installed for the emission of xenon scintillation light by $\alpha$-particles moving through and ionizing the xenon within the cell.

The entire copper cup is enclosed within the inner steel vacuum vessel. The vessel is connected to electrical feedthroughs for signal, bias voltage and the Resistance Temperature Detector (RTD) cables as well as connections to the xenon gas inlet system at the top of the vacuum chamber (9). A turbomolecular pump, realizing a pressure better than \SI{2e-5}{\milli bar}, is also connected from the top.

All cables are PTFE coated to minimize outgassing within the xenon cell. Signals are sent to an oscilloscope via isolated BNC feedthroughs. The temperature is monitored with four \SI{100}{Ohm} Pt RTDs attached to the outside of the copper cup at three different heights as well as directly at the SiPM board to monitor the actual SiPM temperature.

The bottom of the inner vacuum vessel sits on top of a copper cold finger (5) which is thermally linked to a liquid nitrogen dewar outside of the outer vacuum. The cooling power can be fine-tuned via ceramic resistors around the cooling finger. This allows cooling down the entire xenon cell to temperatures at \SI{168}{\kelvin} and guarantees long-term stabilization within \SI{0.1}{\kelvin}. The copper cup allows the fast compensation of any heat flow from the top feedthroughs and direct cooling of the detectors.

\section{Measurements}
\label{sec:measurements}
{An overview of the measurements that were carried out can be found in Table~\ref{tab:measurements}. Overall, three LF version SiPMs of the same wafer were measured, one with the Erlangen (LF E) and two with the Stanford setup (LF S1 and LF S2). In addition, one STD version SiPM (STD S1) and a Hamamatsu PMT were measured with the Stanford setup.}
\begin{table}[t]
\setlength\extrarowheight{1pt}
\centering
\begin{tabular}{L{0.1\columnwidth}C{0.13\columnwidth}C{0.15\columnwidth}C{0.13\columnwidth}C{0.23\columnwidth}}
\toprule
\textbf{Device} & \textbf{$\mathbf{U_{\text{op}}}$ [V]}  & \textbf{$\mathbf{U_{\text{break}}}$ [V]}  & \textbf{Setup}  & \textbf{Temperature [K]}\\
LF E & 33-36.5 & 29.42 & Erlangen & 168 \\
LF S1 & 30-33 & 28.74 & Stanford & 169 \\
LF S2 & 30-34 & 28.83 & Stanford & 169 \\
STD S & 23.5-26 & 22.82 & Stanford & 169 \\
PMT & 1190 & - & Stanford & 293 \\
\bottomrule\\
\end{tabular}
\caption{Overview of the key parameters for the measurements of the various devices.}
\label{tab:measurements}
\end{table}

\subsection{SiPM Dark Measurements}
\label{subsec:sipm_dark}

\begin{figure*}[t]
\centering
\includegraphics[width=0.9\textwidth]{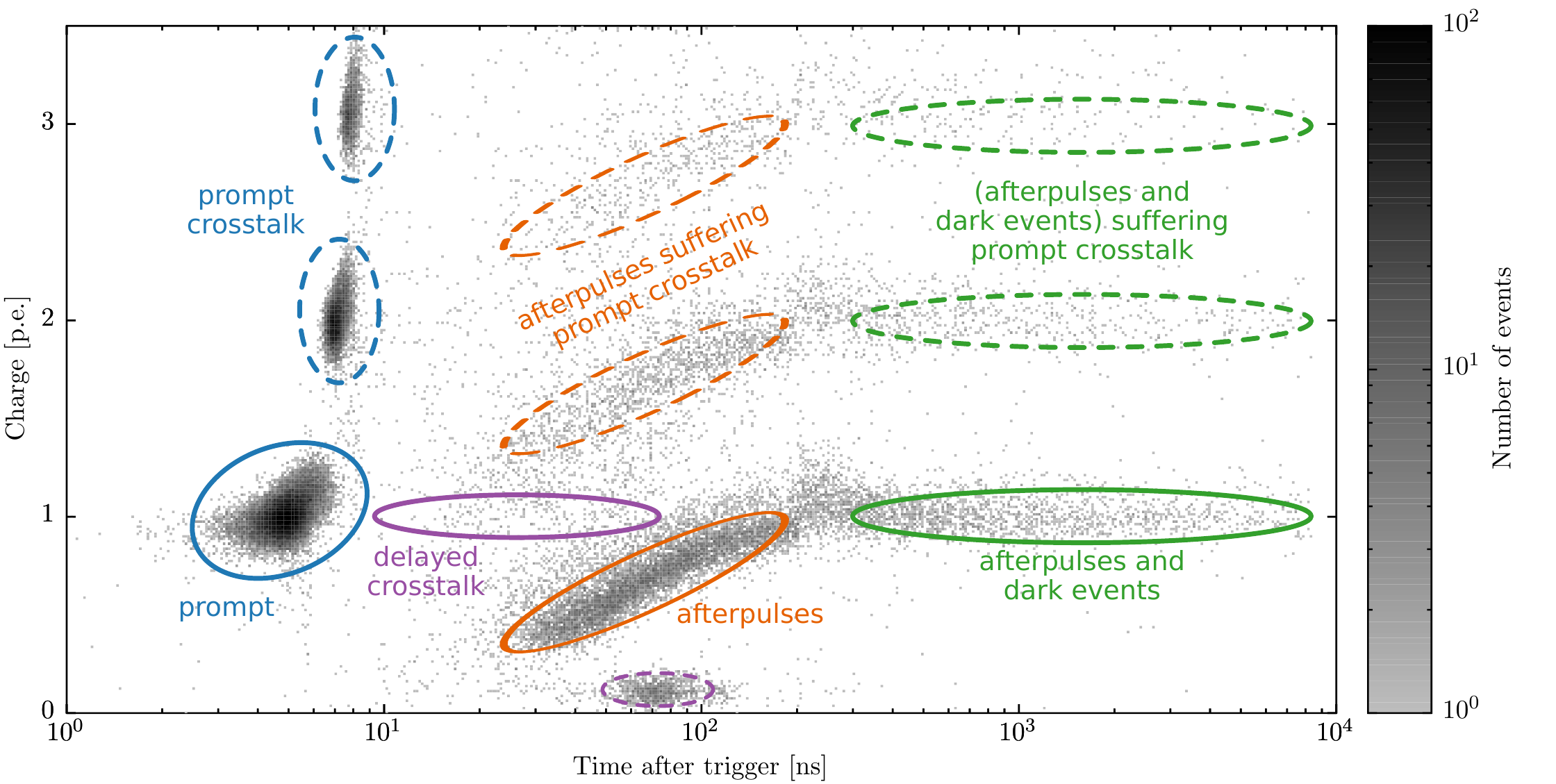}
\caption{Distribution of photoelectron equivalents for parent and subsequent pulses as a function of the time since the microcell was triggered. See text for detailed explanation. Data from Erlangen setup with LF E device at \SI{5.58}{\volt} overvoltage.}
\label{fig:pe_distribution}
\end{figure*}
\begin{figure}[t]
\centering
\includegraphics[width=\columnwidth]{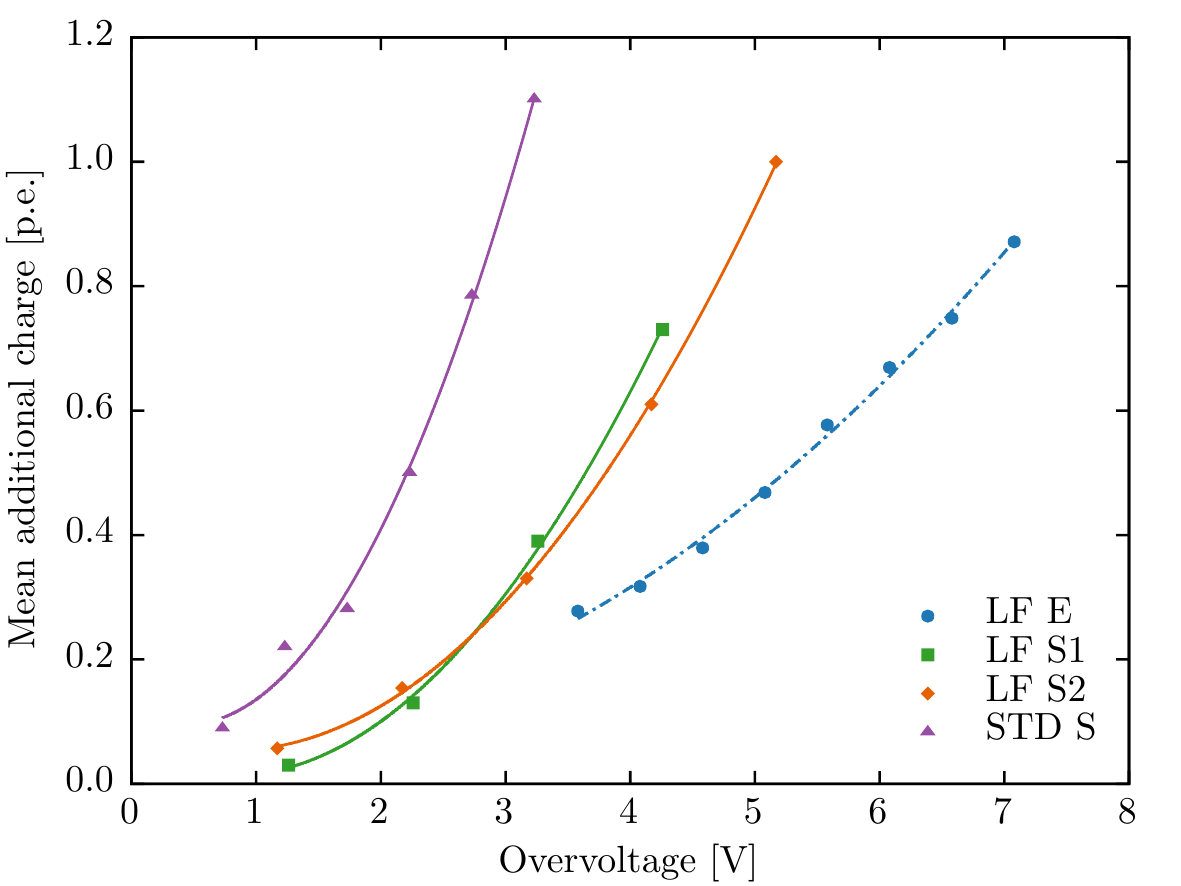}
\caption{The mean contribution of prompt crosstalk {only} (dashed line) and {total} correlated noise (solid lines) as a function of overvoltage, which were measured with the Erlangen (blue) and Stanford (green, orange and purple) setup. {\st{They include correlated avalanche processes to all orders.}} Data from Stanford and Erlangen setup.}
\label{fig:crosstalk}
\end{figure}
Parameters such as dark count rate, recovery time constant, crosstalk and after-pulsing probability were measured in the cryogenic setup in Erlangen at temperatures of \SI{168}{\kelvin}. The xenon chamber was used without a radioactive source and was filled with gaseous nitrogen during these measurements. An FBK LF device (which will be referred to as LF E) was measured between overvoltages of about \SI{3}{\volt} to \SI{7}{\volt}. Data acquisition was triggered by signals exceeding the electronics noise. The waveforms were \SI{10}{\micro\second} long and were sampled with a rate of \SI{2.5}{\giga S\per\second}.

To assess device characteristics, pulses in the waveforms are analyzed in a multi-stage algorithm, starting with the peak finder TSpectrum provided by ROOT~\cite{Root} and using the known pulse shape. $\chi^2$-fits to these pulses are performed with a pulse representation comprising an exponentially modified Gaussian distribution. After a first fit iteration, the pulse shape is set by fixing both rise time and fall time of the pulses to the ensemble's most probable values. {Depending on the overvoltage the rise and fall times would vary between \SI{2.3}{\nano\second} to \SI{2.8}{\nano\second} and \SI{78.7}{\nano\second} to \SI{86.1}{\nano\second}, respectively.} A second fit iteration is performed with fixed pulse shape to improve the estimation of pulse time and amplitude. For fits exceeding a certain threshold of reduced $\chi^2\textsubscript{red}$, test pulses are added iteratively to the fit. The new pulse combination is kept permanently if the value of $\chi^2\textsubscript{red}$ of the new fit improves significantly. Otherwise, the test pulses are discarded. The last step of the algorithm improves the capability to identify overlapping pulses. Using this algorithm, pulses more than about \SI{10}{\nano\second} apart can be separated reliably. The distribution of these amplitudes, converted into number of photoelectron equivalents~(\si{\pe}) as function of the time since the trigger is shown in Figure~\ref{fig:pe_distribution}. This representation allows the identification of the various origins of the background signal, such as thermally induced avalanches. The microcells' recharging process and effects contributing to correlated noise, including prompt and delayed crosstalk and after-pulsing can be identified as well. Prompt pulses (circled in blue) gather around \SI{1}{\pe} and dominantly lie within the first \SI{10}{\nano\second}. Prompt pulses producing prompt crosstalk (circled in dashed blue) are measured as pulses with multiple \si{\pe} as photons emerging from the avalanche are detected quasi-instantaneously in another microcell. Avalanches due to delayed crosstalk (circled in purple), which contributes negligibly, originate from photons being absorbed in passive components of another microcell where the photoelectron diffuses to the active part of that microcell. Due to timing, these delayed avalanches produce another distinct trigger and are identified separately as additional pulses with \SI{1}{\pe}~\cite{Ostrovskiy}. Triggered pulses suffering afterpulsing (circled in orange), describing the release of trapped charge carriers of an avalanche, are measured as a parent pulse and a correlated delayed pulse. Since these delayed pulses can suffer crosstalk as well, the band structure is repeated above \SI{1}{\pe} (circled in dashed orange). Pulses occurring with larger delay to the parent pulse can originate from either afterpulsing or thermal excitation (circled in green) and can suffer prompt crosstalk as well (circled in dashed green). Pulses occurring shortly after a prompt pulse in a previously triggered microcell, which are typically afterpulses, have lower amplitudes since the microcell charge has not been fully replenished yet. The assignment to afterpulsing or {dark events} can only be done on a statistical basis. The pulses marked circled in dashed purple do not correspond to any physical effect but are due to misidentification by the peak finder and are omitted in the further analyses.

\begin{figure}
\centering
\includegraphics[width=\columnwidth]{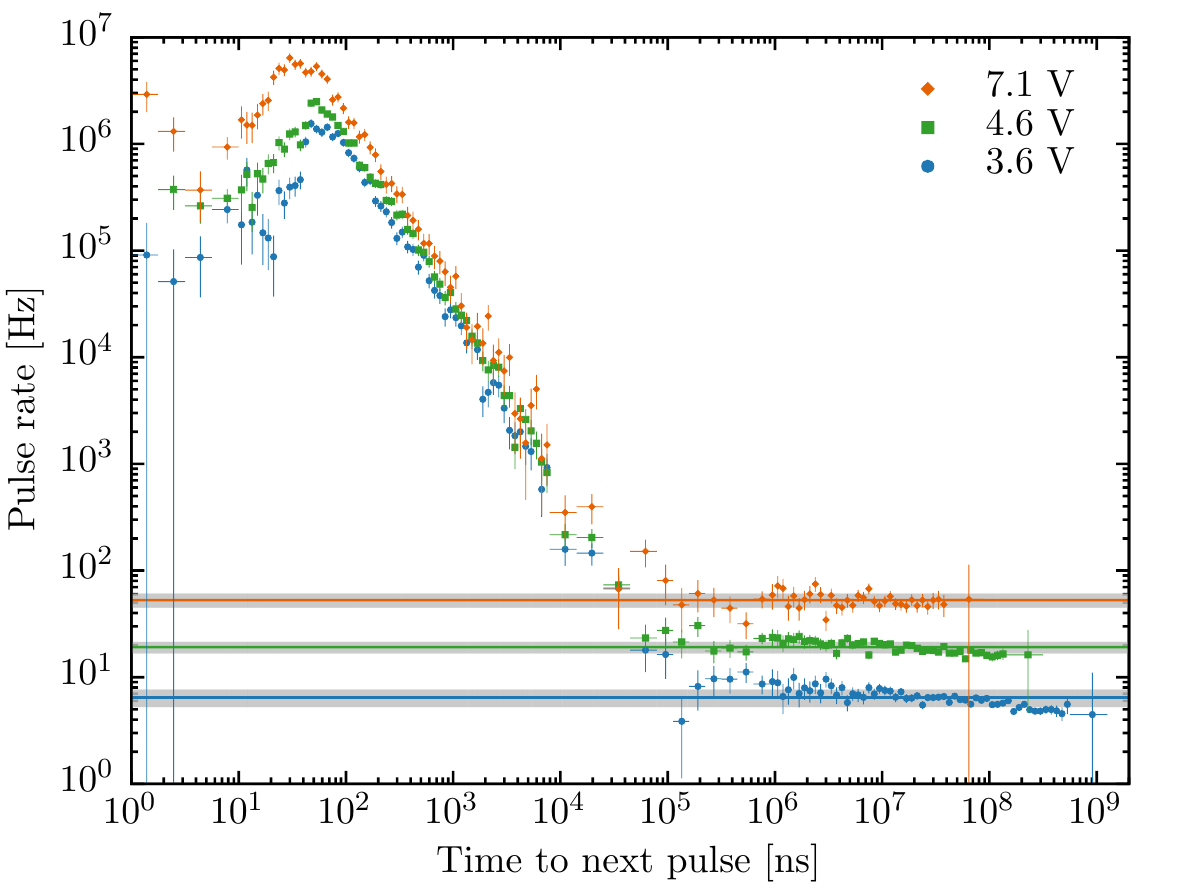}
\caption{Time distribution of the pulse rate $R(t)$ for pulses occurring after a time difference $t$ with respect to the prompt parent pulse. Shown are distributions obtained with an LF device in the Erlangen setup for several overvoltages as denoted in the legend. The error bars are calculated assuming Poissonian uncertainties on the number of events corresponding to each bin. The details on the selection of parent and subsequent pulses are explained in the text. The flat part at long time differences corresponds to the dark count rate whose contributions are indicated by the bands showing the mean value with standard deviation, whereas shorter time differences are dominated by afterpulses. Data from Erlangen setup.}
\label{fig:after_pulsing}
\end{figure}
\begin{figure}
\centering
\includegraphics[width=\columnwidth]{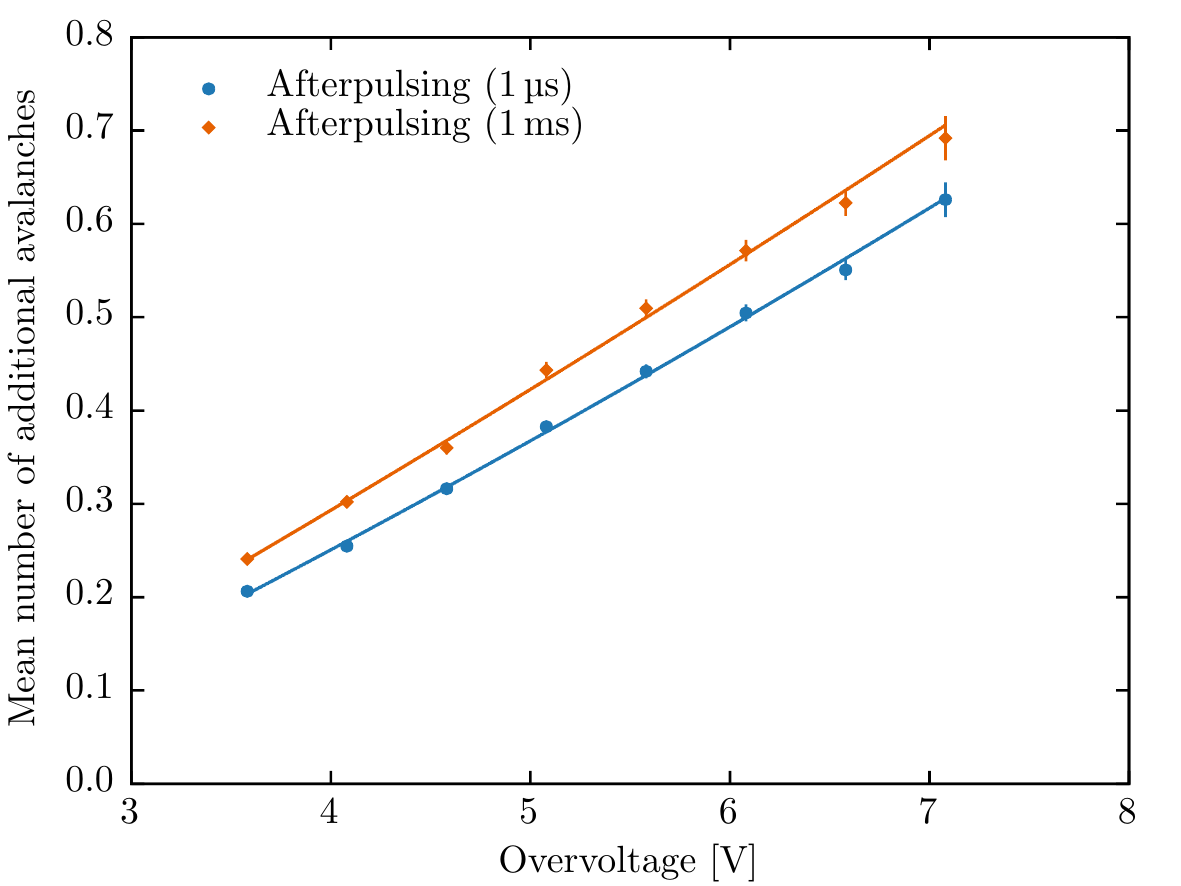}
\caption{The mean number of additional avalanches due to afterpulsing as a function of the overvoltage measured with an LF device in the Erlangen setup. Here, two different time-windows are considered to determine the impact of after-pulsing: the first \SI{1}{\micro\second} and the first \SI{1}{\milli\second} after the primary avalanche. Data from Erlangen setup.}
\label{fig:afterpulsing}
\end{figure}

The charge distribution of prompt pulses is used to determine prompt crosstalk. Due to this process, the average charge in prompt pulses exceeds the mean response of a single microcell $Q_{\SI{1}{\pe}}$. Thus, the mean number of prompt crosstalk avalanches $\overline{N}\textsubscript{CT}$ is conventionally defined as:
\begin{linenomath*}
\begin{equation}\label{eq:sipm_ct}
\overline{N}\textsubscript{CT} =  \frac{1}{N}\sum_{i=1}^N Q_i - Q\textsubscript{\SI{1}{\pe}} \, ,
\end{equation}
\end{linenomath*}
where the normalized sum is the average charge in prompt pulses. This quantity {can be} larger than one and also includes higher order crosstalk processes, i.e.\ avalanches produced by crosstalk that themselves induce additional avalanches via crosstalk. The mean number of prompt crosstalk avalanches $\overline{N}\textsubscript{CT}$ {(blue dashed curve)} is shown in Figure~\ref{fig:crosstalk} for several overvoltages. We emphasize that this quantity is different than the probability for prompt crosstalk to occur.

Afterpulsing is caused by the release of charge carriers trapped within a previously triggered microcell. It can be determined by exploiting the distribution of time differences between parent pulses resulting from a single avalanche and their subsequent pulse. The latter pulse can be of any amplitude exceeding the noise level in order to not reject afterpulses being accompanied by prompt crosstalk. To obtain the {dark count rate} and the mean number of afterpulsing avalanches, the method described in~\cite{Butcher:afterpulsing} is used which introduces a model-independent approach to extract both quantities without the necessity of fitting several time constants for de-trapping in lattice defects.

We emphasize that the mean number of afterpulses does not equal its correlated charge contribution since the released charge of an avalanche depends on the microcell's recovery state. Detailed pulse selection on both primary and secondary pulses are applied as stated in~\cite{Butcher:afterpulsing}. Assuming Poissonian statistics, {the probability $p_i$ for a secondary pulse to occur between times $t_i$ and $t_{i+1}$ relative to the initial pulse is given by:
\begin{linenomath*}
\begin{equation}
p_i = e^{-\beta_i}\cdot(1-e^{-\lambda_i})
\label{eq:ap_probability}
\end{equation}
\end{linenomath*}
where $\beta_i$ and $\lambda_i$ denote the average number of correlated pulses before time $t_i$ and between time $t_i$ and $t_{i+1}$, respectively. The first part of the function accounts for no pulses up to time $t_i$, whereas the second part accounts for at least one pulse occurring between time $t_i$ and $t_{i+1}$. Exploiting the sequential nature of $\beta_i=\sum_{j=0}^{i-1}\lambda_j$, one can calculate $\lambda_i$ iteratively for any time window by inverting Equation~\ref{eq:ap_probability} and using $\beta_0=0$. The correlated pulse rate can then be simply calculated as
\begin{linenomath*}
\begin{equation}
R_i = \frac{\lambda_i}{t_{i+1}-t_{i}}\, .
\end{equation}
\end{linenomath*}}

\begin{figure}[t]
\centering
\includegraphics[width=\columnwidth]{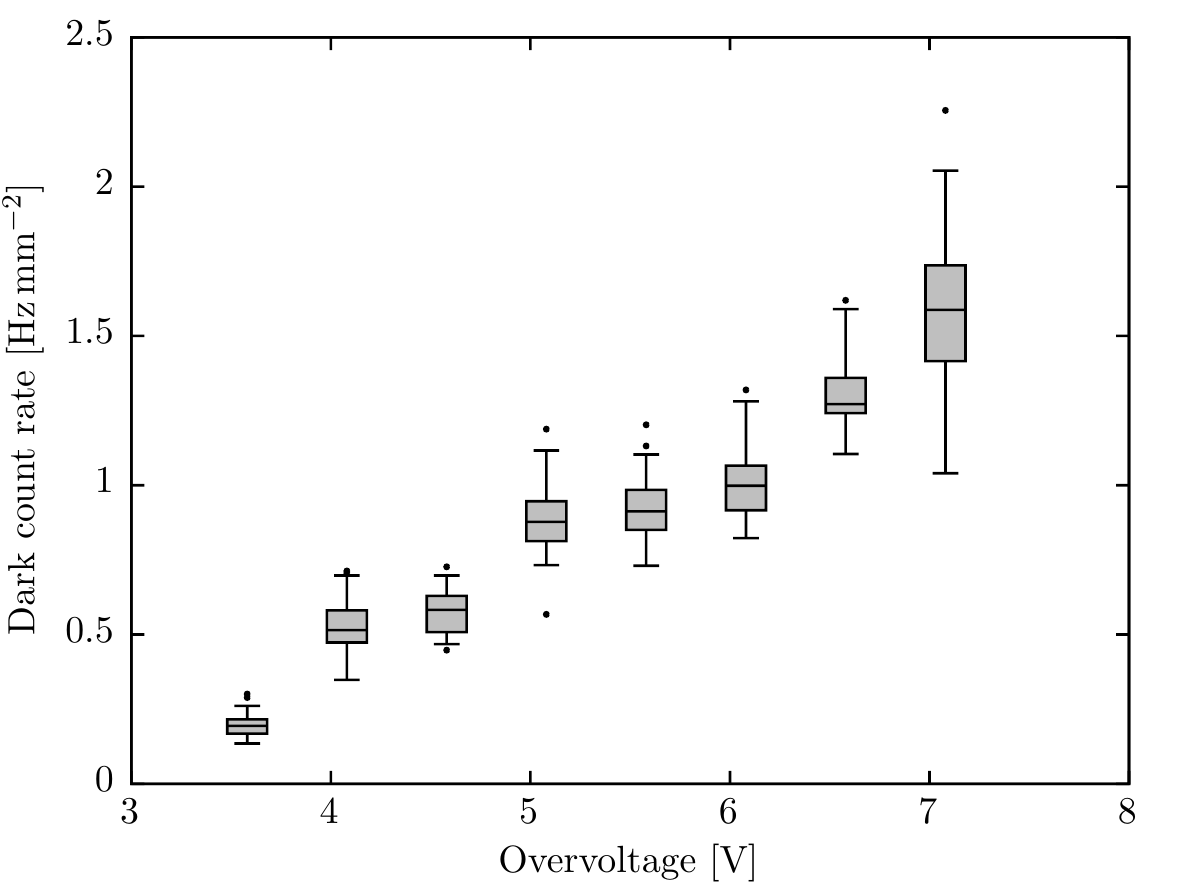}
\caption{Box plot of {dark count rate} per unit area of the LF E device for several overvoltages. The surface area on the front face of the device is $\num{5.56}\times\SI{5.96}{\milli\metre\squared}$. {Data points with delays exceeding \SI{1}{\milli\second} in the distributions in Figure 8 contribute to this plot. The box indicates the area between the 1st and 3rd quartile (\SI{25}{\percent}-\SI{75}{\percent} of the data points) whereas the 2nd quartile (median value) is shown as horizontal bar inside the box. The whiskers comprise \SI{95}{\percent} of the data points around the median. Data points outside that range are shown individually as outliers}. Data from Erlangen setup.}
\label{fig:dcr_linear}
\end{figure}

Example distributions of the pulse rate are shown in Figure~\ref{fig:after_pulsing}. The excess in rate due to afterpulses vanishes for long time differences, leaving only the contribution due to {dark events}. Summing the excess at time differences below \SI{1}{\micro\second} yields the mean number of afterpulses that is of particular interest for nEXO and is shown in Figure~\ref{fig:afterpulsing} for several overvoltages. Assuming no contributions from afterpulsing after \SI{1}{\milli\second}, we can consider this number to be the overall mean number of afterpulses. The {dark count rate} can be extracted by fitting the part of the distribution above \SI{1}{\milli\second} with a flat function. Even though the distributions in our measurements are not perfectly flat after \SI{1}{\milli\second}, we use them to determine the {dark count rate}. For this reason, we state the distribution of values at a given overvoltage rather than the mean value as shown in Figure~\ref{fig:dcr_linear}. The {dark count rate} does not exceed \SI{2}{\hertz\per\milli\meter\squared}, well below the limit of \SI{50}{\hertz\per\milli\meter\squared} required by nEXO. The contribution due to delayed crosstalk is not modeled separately here but is incorporated in the mean number of afterpulses. Its contribution compared to afterpulsing is negligible as can be seen in Figure~\ref{fig:pe_distribution}.

\subsection{Reference PMT Calibration}
\begin{figure}[t]
\centering
\includegraphics[width=\columnwidth]{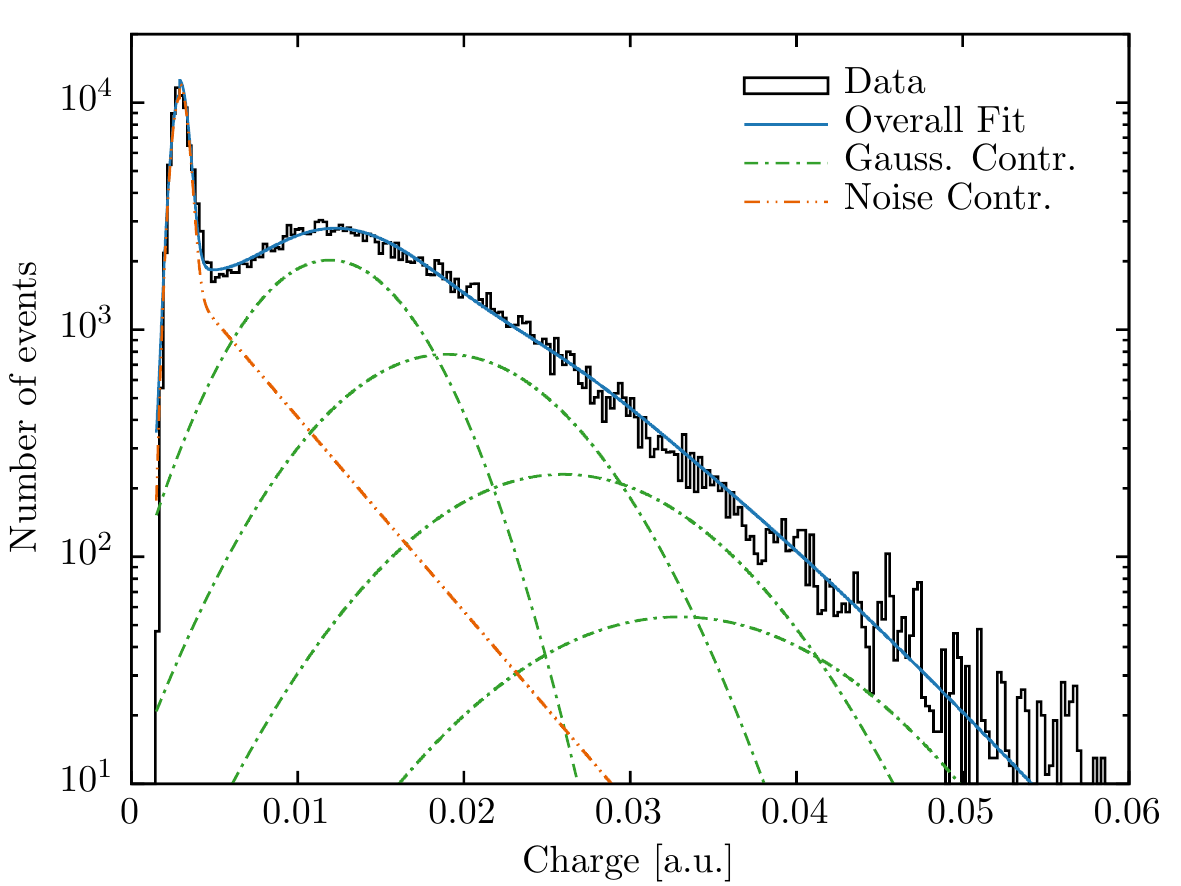}
\caption{Typical response of the reference PMT at Stanford to an LED operated at low light level. By fitting the function described in Formula~\ref{eq:pmt_1pe}, the value for the \SI{1}{\pe} charge response $Q_1$ can be extracted and used to calibrate the PMT. The overall fit is shown as solid blue line, whereas the different \si{\pe} orders of the fit are dashed green and the background contribution is shown as a dashed orange curve. Data from Stanford setup.}
\label{fig:pmt_1pe_response}
\end{figure}

In order to measure the absolute PDE of the SiPMs reliably, a calibrated R9875P PMT from Hamamatsu~\cite{pmt} was used as a reference detector in the Stanford setup. The PMT has a \SI{11}{\milli\meter} wide synthetic silica entrance window and Cs-Te photocathode with roughly \SI{8}{\milli\meter} in diameter. It was operated at \SI{1190}{\volt} for which a quantum efficiency (QE) at \SI{175}{\nano\meter} of \SI{14}{\percent} and a collection efficiency (CE) of the first dynode of \SI{70.6}{\percent} was measured by Hamamatsu.

The single photon response can be extracted from a measurement with a pulsed LED. The PMT is placed inside the detector cage and the xenon scintillation light was blocked by placing a thick black sheet on top of the bandpass filter. The light level of the LED inside the vacuum chamber is set very low, such that the PMT roughly only sees about one photon every ten LED pulses. An ideal photomultiplier response follows a Poisson distribution. However, there are various background processes that will contribute to the signal, e.g.\ thermoelectric emission from the photo-cathode or the dynodes, and ultimately lead to an additional charge noise contribution. The contribution of these processes to the charge spectrum can be expected to decrease exponentially with charge. Thus, the real PMT response will be composed of a Poisson-convoluted Gaussian distribution of the true signal and the distribution of the noise contribution. It can be expressed as~\cite{bellamy}
\begin{linenomath*}
\begin{align}
& f(q) =  \left[\frac{1-\omega}{\sigma_0 \sqrt{2\pi}}\cdot\exp\left(-\frac{(q-Q_0)^2}{2\sigma_0^2} \right ) \right.  \\
& + \vspace{-15pt}\left. \frac{}{}\hspace{-2pt}\omega\cdot \theta(q-Q_0)\cdot\alpha\exp\left(-\alpha(q-Q_0) \right ) \right] \cdot e^{-\mu}  \nonumber\\
&+ \sum_{n=1}^k \frac{\mu^ne^{-\mu}}{n!}\frac{1}{\sigma_1\sqrt{2\pi n}} \exp\left(-\frac{(q-Q_0-Q_{\rm sh}-nQ_1)^2}{2n\sigma_1^2} \right) \nonumber \label{eq:pmt_1pe}
\end{align}
\end{linenomath*}
where $q$ is the measured charge, $\omega$ is the probability for a background process to happen, $\alpha$ the coefficient of exponential decrease of discrete background processes and $\mu$ the mean number of the detected {photoelectrons}. $Q_0$, $\sigma_0$, $Q_1$ and $\sigma_1$ are the mean value and the standard deviation of the pedestal and the single photon response, respectively. Furthermore, $Q_{\rm sh} = \omega/\alpha$ introduces a shift of the true \si{\pe} pulses due to the background contribution. An example fit to the PMT response is shown in Figure~\ref{fig:pmt_1pe_response}.

\subsection{Absolute PDE Measurement}
\label{subsec:pde}
The absolute PDE of two LF (LF S1 and LF S2) and one STD (STD S) version SiPMs was measured with the Stanford setup. The PDE determination is based on a comparison of the amount of light that was seen by both the calibration PMT and the SiPM. Given the known QE and CE of the PMT and a proper calibration of the single photon response, one can determine the amount of light detected by the reference detector. Together with the measured response of the SiPMs one can translate this into an absolute PDE for the SiPM.

At Stanford the SiPM signals are transported with flexible solid-core cables and through a grounded-shield BNC feedthrough. A CAEN digitizer (DT5724~\cite{caen}) records \SI{10}{\micro\second} long waveforms with a sampling rate of \SI{100}{\mega s\per\second} after the signal has been amplified in a Cremat charge-sensitive amplifier (CR-110~\cite{cremat_amp}) and a Gaussian shaper (CR-200~\cite{cremat_shaper}) with a \SI{100}{\nano \second} shaping time. The data is analyzed with a custom C++-based software that uses ROOT~\cite{Root}. The waveforms were integrated over a fixed time window of \SI{1}{\micro \second} after the trigger to obtain the collected charge of the event. To allow a comparison of the mean light fluence seen by both detectors, the integrated charge was normalized by the respective gain and the surface area of the devices.

\begin{figure}[t]
\centering
\includegraphics[width=\columnwidth]{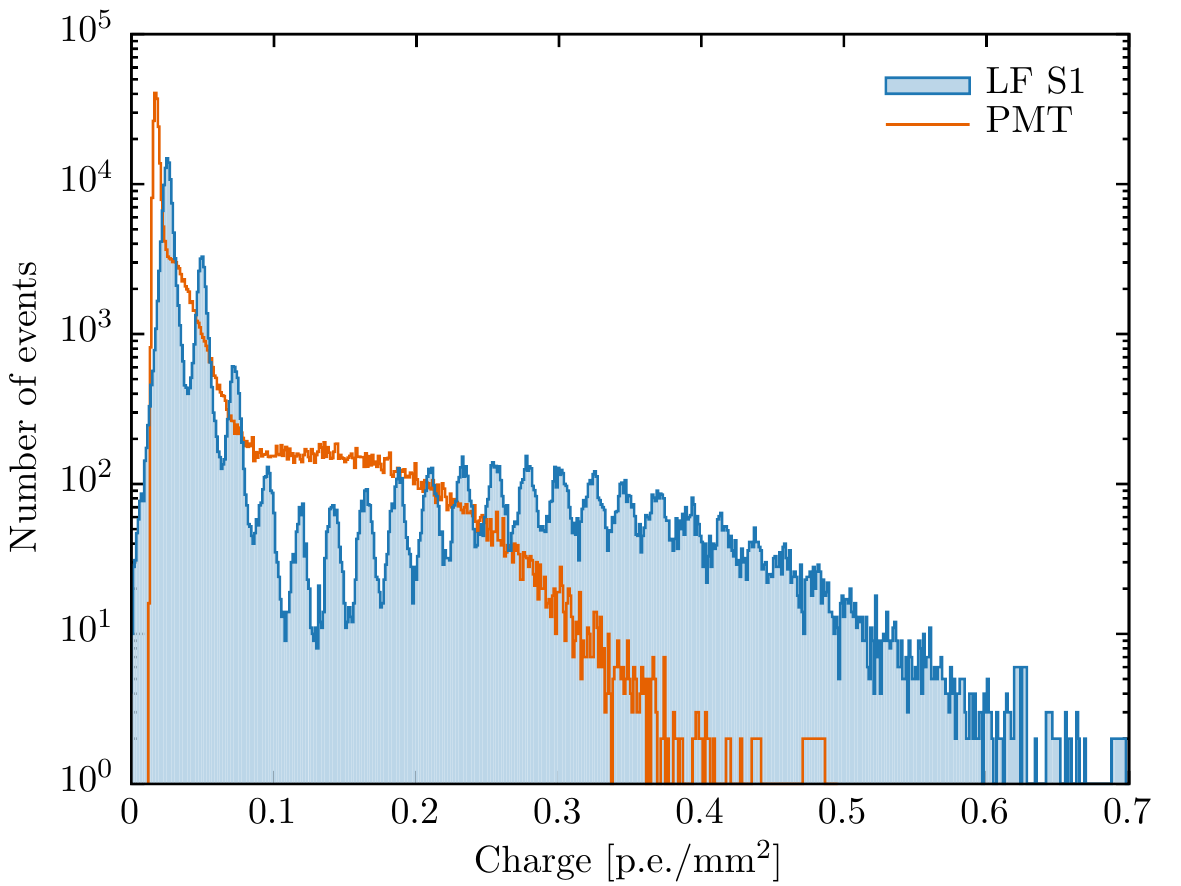}
\caption{Example charge spectrum of {xenon scintillation light from} the $^{252}$Cf source taken with a LF device (blue) at \SI{1}{\volt} over-voltage and with the reference PMT (orange). The $x$-axis is normalized by the gain, the surface area of the devices and in the case of the SiPM by the correlated noise contribution. The left parts of the spectrum corresponds to {dark events} and alpha particles. The broader distributions in both spectra are due to spontaneous fission events. It is centered around \SI{0.23}{\pe/\milli\meter\squared} and \SI{0.37}{\pe/\milli\meter\squared} for the PMT and SiPM, respectively. Data from Stanford setup.}
\label{fig:sipm_spectrum}
\end{figure}

Figure~\ref{fig:sipm_spectrum} shows the spectrum of the $^{252}$Cf source that was measured with the PMT and the SiPM. The abscissa shows the collected charge as {photoelectron} equivalents per unit area (\si{\pe/\milli\meter\squared}). There are three key features in the spectrum obtained with the PMT (orange). The first one is the very narrow peak at small values. This sharp pedestal peak is followed by a second peak that contains events originating from an $\alpha$-decay (\si{\mega\electronvolt} energies). The more distinct and broader distribution centered at about \SI{0.23}{\pe/\milli\meter\squared} can be assigned to fission fragments after the spontaneous fission of $^{252}$Cf, which typically releases energy in the order of tens of \si{\mega\electronvolt} and therefore has a larger light output. In the case of the SiPM (blue), here a LF device, we do not observe a pedestal peak, but see the single photon resolving capabilities of these devices. The first few peaks correspond to {dark events} and $\alpha$-particle events, whereas the broad distribution centered around \SI{0.37}{\pe\per\milli\meter\squared} can be assigned to fission fragments. A combination of an exponentially decreasing background and a Gaussian is fitted to extract the position of the fission peak.

The correlated noise probability of the SiPM was measured with the same setup, blocking the Xe scintillation light by placing a thick black sheet on top of the bandpass filter and recording {dark events}. The mean of the pulse height spectrum, in the absence of correlated noise, should be exactly \num{1}. However, photon-triggered avalanches can be accompanied by correlated avalanches, as described earlier, and therefore the mean of the pulse height spectrum deviates from unity. Thus, the excess to unity represents the mean number of additional avalanches per parent avalanches within \SI{1}{\micro\second}. This includes all higher order correlated avalanche processes, such as avalanches due to crosstalk which themselves suffer afterpulsing and vice versa.

\begin{figure*}
\centering
\includegraphics[width=\columnwidth]{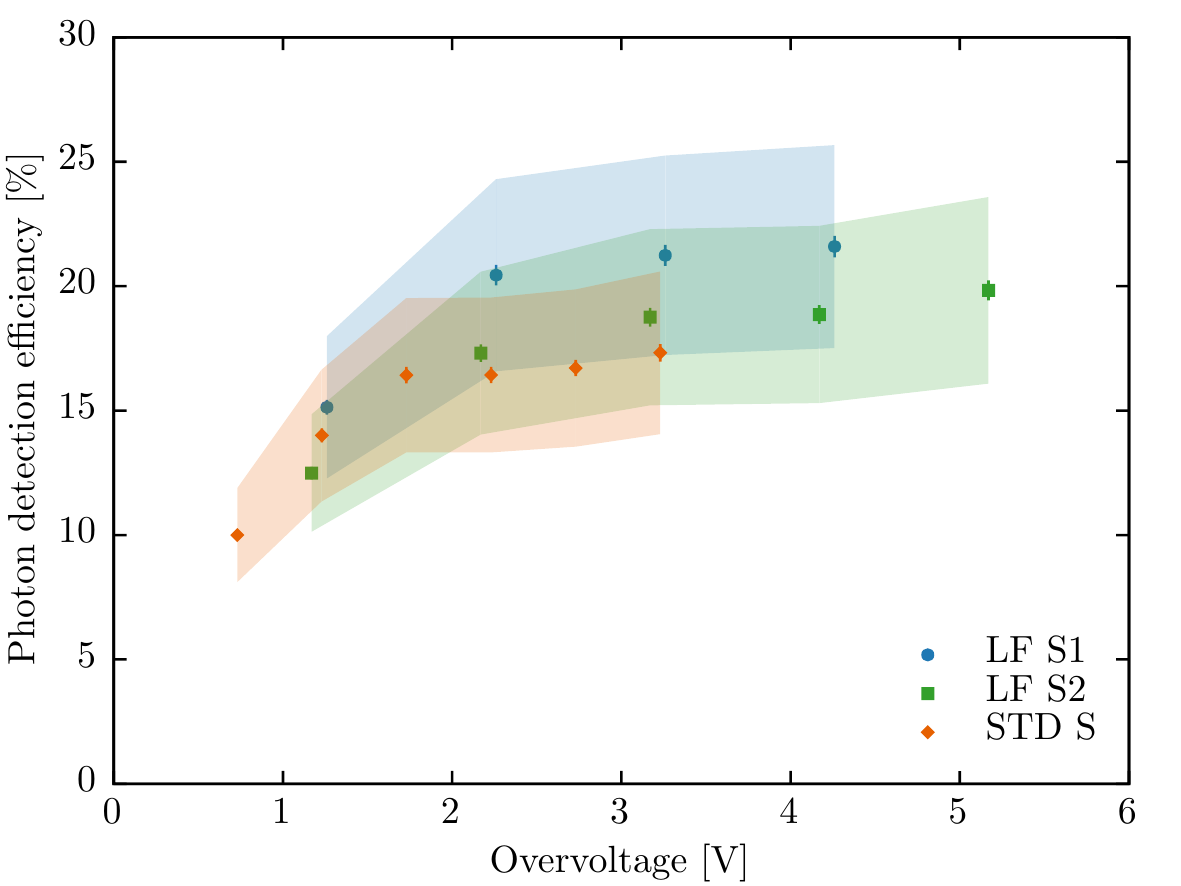}
\includegraphics[width=\columnwidth]{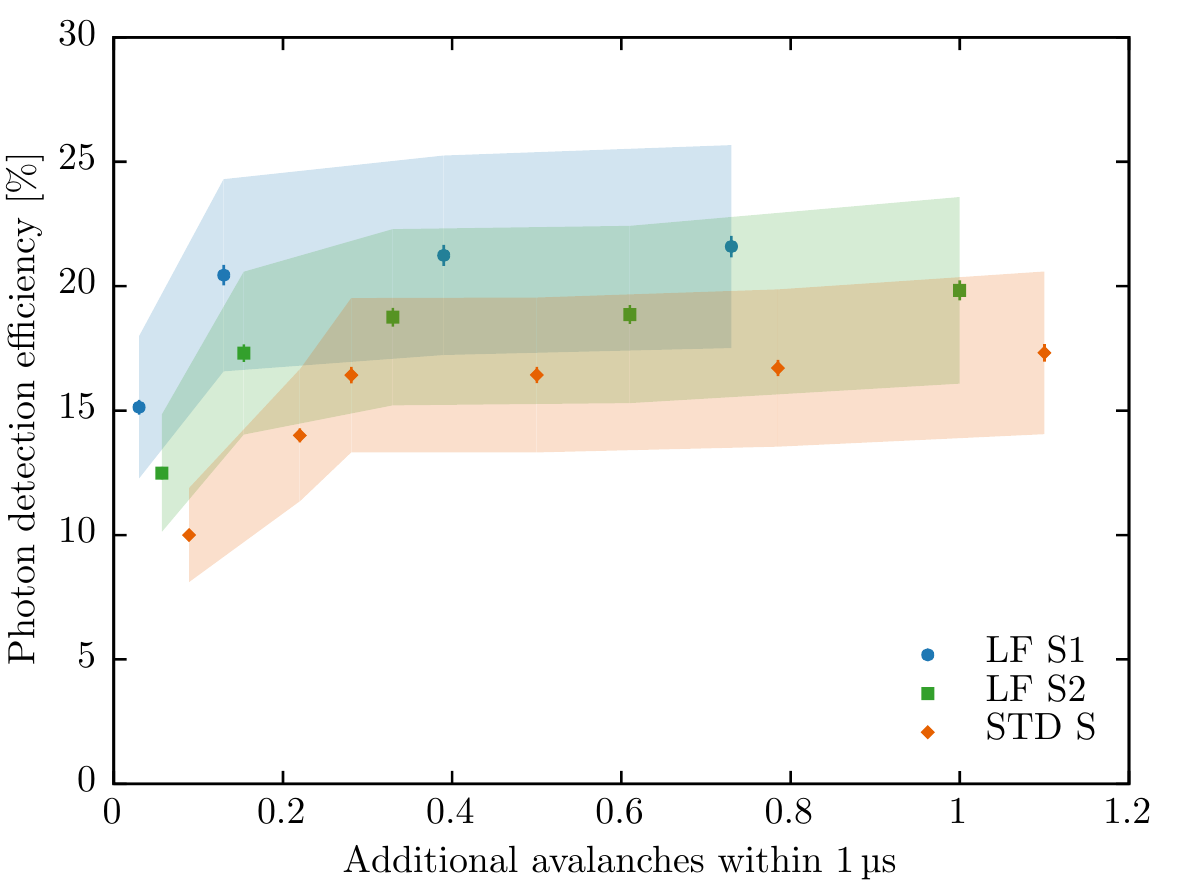}
\caption{Photon detection efficiency as a function of overvoltage (left) and the number of additional correlated avalanches within \SI{1}{\micro\second} (right). The error bars represent the statistical errors whereas the colored bands show the systematic errors, that are stated in Table~\ref{tab:errors}. These measurements show that within the uncertainties the required \SI{15}{\percent} PDE, while still having a correlated avalanche probability lower than \SI{20}{\percent}, are fulfilled for nEXO. However, a very small range of operation, only up to a few volts overvoltage, is possible. Data from Stanford setup.}
\label{fig:pde_measurements}
\end{figure*}

Given the ratio of the position of the fission peaks for both detectors and the known total efficiency of the PMT of {\SI{9.9}{\percent}}, an absolute PDE value for the SiPM can be extracted. The results for three different (two LF and one STD type) devices are shown in Figure~\ref{fig:pde_measurements}. The plot on the left-hand side shows the dependence of the PDE with over-voltage, where a slight increase is observable. However, the efficiency saturates at relatively low bias voltages, corresponding to a saturation of the probability to trigger an avalanche. The plot on the right-hand side shows the PDE as a function of the number of additional avalanches within \SI{1}{\micro\second}. It shows that the LF devices meet our requirements of more than \SI{15}{\percent} PDE while keeping the correlated noise probability below \SI{20}{\percent}. Generally, the LF devices seem to have a better operating range and performance since for a given correlated noise probability they can be operated at a higher overvoltage and therefore have a higher PDE. Also, from an electronics point of view a higher overvoltage is desirable since this automatically means higher gain and better signal-to-noise ratio.

\subsection{Systematic and Statistical Errors}
\begin{table}[t]
\setlength\extrarowheight{1pt}
\centering
\begin{tabular}{L{0.6\columnwidth}R{0.3\columnwidth}}
\toprule
\textbf{Source of Systematic Error } & \textbf{Value [\si{\percent}]} \\
Solid angle & 3 \\
Angular Distribution & 1 \\
PMT gain stability & 6 \\
PMT gain modeling & 10 \\
PMT QE uncertainty & {$\sim$ 2} \\
PMT CE uncertainty & {$\sim$ 14} \\
\textbf{Quadratic sum} & {$\mathbf{\sim 19}$} \\
\midrule
\textbf{Source of Statistical Error } & \textbf{Value [\si{\percent}]} \\
PMT gain calibration & 1 \\
SiPM gain calibration & 1 \\
Correlated noise correction & 1 \\
Fission peak position & 1 \\
\textbf{Quadratic sum} & $\mathbf{2}$ \\
\bottomrule\\
\end{tabular}
\caption{Summary of systematic and statistical error sources. The latter one is the standard deviation of the corresponding fit parameter. The largest component of the uncertainty in measuring the PDE is the uncertainty on the CE of the PMT. Since the error sources are independent, the components are added in quadrature.}
\label{tab:errors}
\end{table}
The colored error bands and error bars in Figure~\ref{fig:pde_measurements} stem from several sources of systematic and statistical uncertainties, respectively. The individual components and their estimated values are summarized in Table~\ref{tab:errors}. The PMT introduces the major components for the systematics. The reference PMT has been calibrated by Hamamatsu in order to get the respective wavelength-dependent QE and the CE of this particular tube. {The calibration was done at room temperature and at normal incidence}. The QE was measured to be \SI{14}{\percent} with a relative uncertainty of {about \SI{2}{\percent}}. For the collection efficiency (CE) at the first dynode of the PMT a value of \SI{70.6}{\percent} was determined by Hamamatsu without any estimate for the uncertainty. Conservatively, we assume $(70\pm 10)\si{\percent}$ for the CE of the PMT. Furthermore, a variation of the PMT gain over time on the order of \SI{6}{\percent} was observed, which is due to temperature related gain variations over time, caused by not cooling the PMT during the measurements.
The statistical uncertainty on the PMT gain itself is relatively small and well below \SI{1}{\percent}. In order to cover possible variations due to the model dependence of our gain calibration we assign another \SI{10}{\percent} as an systematic error for the gain calibration. The bottom inset of the copper box is designed such that all detectors are at the same distance relative to the light source. In order to take $\num{1} - \SI{2}{\milli\meter}$ offset variations with respect to the light source at a distance of roughly \SI{108}{\milli\meter} and different detector geometries into account, a \SI{3}{\percent} systematic error is considered for the solid angle. The value was extracted from an optical simulation of the setup with different detectors, using the Chroma software~\cite{chroma}. Furthermore, the simulation predicts that a large fraction of photons are detected without having been reflected and at angles close to normal incidence, i.e.\ $\ang{3\pm 1}$.
From our simulation we can infer that the effect of non-normal incidence on the PDE from photons that are reflected off the PTFE sphere is about \SI{\sim 1}{\percent}.
The SiPM gain calibration can be performed reliably and with a sub-percent level accuracy, which is why we assign \SI{1}{\percent} statistical error. Similarly for the position of the fission peak, which is extracted by fitting a combination of an exponential and a Gaussian function. We assign a value of \SI{1}{\percent} for this uncertainty as well.
Finally, the uncertainty on the correlated noise was determined as the standard error on the mean of the pulse distribution that is obtained during a dark measurement and was found to be \SI{1}{\percent}.

\section{Expected Performance for nEXO}
\label{sec:nexo_performance}
The energy resolution in nEXO can be evaluated using a simple model that is useful to understand the trade-off between PDE and correlated noise that both increase with over-voltage. Two quantities are commonly measured in a LXe TPC in order to optimize the energy resolution~\cite{Conti}: the scintillation light and the number of released electrons which did not recombine. In the nEXO TPC, the electrons that are released during ionization of the Xe atoms and did not recombine drift towards the anodes. We assume that due to the anti-correlation, each ionization electron that recombines produces a photon. The fraction of recombining electrons will be denoted as $R$. Hence, event by event fluctuations of the number of electrons $Q$ and photons $S$ produced are canceled out by performing the optimal linear combination of the two channels.

The production of ionization electrons $Q$ and scintillation photons $S$ can be parameterized as
\begin{linenomath*}
\begin{align}
Q & =  \frac{E}{W} \cdot (1-R) \\
S & =  \frac{E}{W} \cdot (S_i+R)
\end{align}
\end{linenomath*}
with $E$ the deposited energy (\SI{2458.07\pm 0.31}{\kilo\electronvolt} during the $0\nu\beta\beta$ decay of $^{136}\mathrm{Xe}$~\cite{qvalue,qvalue2}), $W$ the effective energy required to create an electron-ion pair ($W=\SI{15.6}{\electronvolt}$~\cite{doke,nest}) and $S_i$ the fraction of scintillation photons produced by excitation relative to ionization. $S_i$ is set to the most probable value of \num{0.13} according to~\cite{Aprile}.
The optimum energy estimator $O$ canceling the fluctuation of $R$ is
\begin{linenomath*}
\begin{equation}
O = S + Q = \frac{E}{W} \cdot (S_i+1) \, .
\label{eq:optest}
\end{equation}
\end{linenomath*}
under the assumption that the electronics noise is smaller than the recombination noise.

In nEXO, the energy resolution is dominated by the noise in the light channel. Therefore, good energy resolution is strongly coupled to the overall light collection efficiency. The scintillation light will be detected by SiPMs covering up to $\SI{4}{\meter\squared}$. Moreover, the transport and photon detection efficiency are entangled because the SiPM surface is very reflective at {\SI{175}{\nano\meter}} due to the large differences of the indices of refraction between silicon ($n_{\rm Si}=\num{0.682}$), silicon dioxide ($n_{{\rm SiO}_2}=\num{1.61}$) and LXe ($n_{\rm LXe} = \num{1.66}$). {The VUV-HD generation SiPMs have a \SI{1.5}{\micro\meter} thick ${\rm SiO}_2$ layer on the top surface as an anti-reflective coating}. Assuming a PDE of \SI{15}{\percent}, our simulations show that an $\epsilon_o$, i.e.\ the product of the PTE and the SiPM PDE, of $\geq\SI{3}{\percent}$ is sufficient to achieve an energy resolution of $\sigma/Q_{\beta\beta} \leq \SI{1}{\percent}$. This corresponds to a PTE of \SI{20}{\percent}, which can already be achieved by a \SI{60}{\percent} reflectivity of the cathode and the field shaping rings and a \SI{50}{\percent} reflectivity of the anode. We would like to emphasize that these assumptions are relatively conservative.
Therefore, the average number of photo-electrons is then \num{1985} at \SI{2458}{\kilo\electronvolt}. The average number of {dark events} within the longest possible integration window of \SI{1}{\micro\second} and for the maximum measured value of \SI{2}{\hertz\per\milli\meter\squared} is only \num{10} and is therefore negligible.

As for the noise in the charge channel, ionization electrons are detected on pads without any amplification besides the electronics preamplifier stage that introduces an equivalent noise charge of about $\sigma_q = \SI{200}{e^-}$ per channel. Given the current choice of \SI{3}{\milli\meter} pixel pitch (studies of the optimal value are still ongoing), the charge of a $0\nu\beta\beta$ decay will be distributed over about 10 channels on average, which will bring the equivalent noise charge back to about $\sigma_q = \SI{600}{e^-}$. This is comparable to the value that was achieved by \mbox{EXO-200} for one channel ($\sigma_q = \SI{800}{e^-}$) with a readout pitch of \SI{9}{\milli\meter}. We can consider this to be roughly the cut-off value below which the charge noise can be considered sub-dominant for the overall energy resolution, as indicated by simulations where the effects of diffusion, electronics noise, and the channel multiplicity are taken into account. This holds true as long as the main contribution to the energy resolution is noise from the light channel. As we improve $\epsilon_0$ the improvement of the energy resolution will then be limited by $\sigma_q$.
Other sources of fluctuations are known to be negligible.

With these assumptions, the average number of detected SiPM avalanches $A$ can be written as
\begin{linenomath*}
\begin{equation}
A =\epsilon_o \cdot  S \cdot (1+\Lambda)
\end{equation}
\end{linenomath*}
with $\epsilon_o$ the overall efficiency of detecting a scintillation photon, $S$ the average number of scintillation photons, and $\Lambda$ the average total number of correlated avalanches per avalanche within \SI{1}{\micro\second} at all orders.
The optimum estimator can then be written as
\begin{linenomath*}
\begin{equation}
O = Q + \frac{A}{\epsilon_o (1+\Lambda)} \, .
\end{equation}
\end{linenomath*}
Introducing the Fano factor $F_{QS}$ for the combined production of scintillation and ionization, the energy resolution (fluctuations of the optimum estimator) can be quantified as
\begin{linenomath*}
\begin{equation}
\sigma_O^2 = F_{QS} \cdot O + \sigma_Q^2 +  \frac{S}{\epsilon_0} \left[ (1-\epsilon_0) + \frac{\Lambda}{(1+\Lambda)^2} \right] \label{eq:resolution}
\end{equation}
\end{linenomath*}
in the limit of small $\epsilon_0$. The value of $F_{QS}$ is not well known, and it is set to 1 conservatively although it could be as low as \num{0.1}~\cite{doke}. However, its contribution compared to the other factors is completely negligible. This equation also assumes that the light detection is dominated by binomial fluctuations due to the finite efficiency of the SiPMs. Fluctuations of the production of correlated avalanches are modeled using Poisson statistics for simplicity. Gain fluctuations and electronics noise for the light channels can be safely neglected.

\begin{figure}[t]
\centering
\includegraphics[width=\columnwidth]{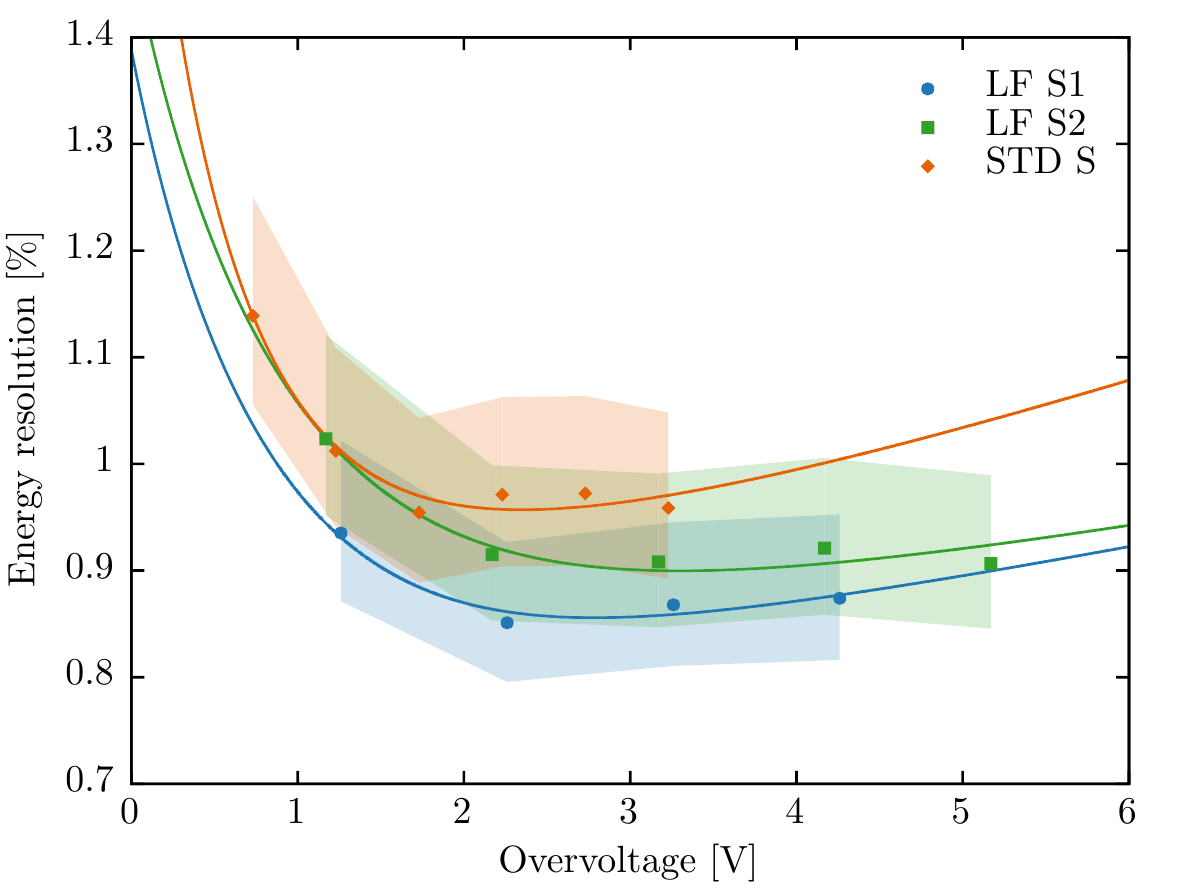}
\caption{Estimated energy resolution {(as defined in Equation~\ref{eq:resolution})} for nEXO as a function of the overvoltage at which the SiPMs are operated. Using a simple model, that is described in more detail in the text, the optimal operation voltage for the SiPM can be determined in order to minimize the energy resolution for nEXO. Due to the trade-off between increasing PDE and correlated noise, the minimal energy resolution is in general not achieved by choosing the largest overvoltage. The colored bands are propagated based on the systematic errors on the PDE. The curves are solely added as visual aid. Data from Stanford setup.}
\label{fig:resolution_vs_efficiency}
\end{figure}

By using the values of $\epsilon_0$ and $\Lambda$ obtained in the previous sections, this simple model allows the derivation of a functional dependence of the energy resolution on the overvoltage of the SiPMs, as shown in Figure~\ref{fig:resolution_vs_efficiency}. The most dominant feature of this function is the existence of a minimum. Furthermore, the minimum does not coincide with the largest operation overvoltage, which in general results in a worse energy resolution because of the increase of correlated noise, mostly due to afterpulsing.

\section{Conclusion}
\label{sec:conclusion}
A study of recent VUV-sensitive FBK SiPM types was presented with a promising SiPM candidate that would meet nEXO's requirements on the photon detection efficiency and the probability for correlated avalanches within the uncertainties. Most specifically, the LF devices exceed the necessary \SI{15}{\percent} PDE while maintaining a correlated noise probability of less than \SI{20}{\percent}. {More specifically, at about \SI{2.2}{\volt} overvoltage the LF S1 and LF S2 devices have a PDE of about \SI{20}{\percent} and \SI{17}{\percent} and a CN of \SI{13}{\percent} and \SI{15}{\percent}, respectively.}
While these measurements were carried out in vacuum and simulations suggest a straightforward translations of these results into LXe, it remains to be confirmed by measurements in LXe. Using a simple model, we estimate the energy resolution of nEXO, taking into account the competing effects of increasing PDE and correlated noise with overvoltage. We predict to be able to reach and possible surpass the anticipated $\sigma/Q_{\beta\beta} = \SI{1}{\percent}$.

Looking forward, work is in progress to build a large scale light detector module using the LF SiPMs, to be tested in LXe. Furthermore, work on readout electronics for a large scale SiPM module is currently in progress with the goal to have a light detection module solution for nEXO. In addition, a parallel effort will study the angular dependence of the PDE in LXe, which was estimated to be negligible in the setup used for this study compared to the total systematic error budget.

\section*{Acknowledgment}
This work has been supported by the Offices of Nuclear and High Energy Physics within DOE’s Office of Science, and NSF in the United States, by NSERC, CFI, FRQNT, NRC, and the McDonald Institute (CFREF) in Canada, by BaCaTeC in Germany, by SNF in Switzerland, by IBS in Korea, by RFBR in Russia, and by CAS and ISTCP in China. This work was supported in part by Laboratory Directed Research and Development (LDRD) programs at Brookhaven National Laboratory (BNL), Lawrence Livermore National Laboratory (LLNL), Oak Ridge National Laboratory (ORNL) and Pacific Northwest National Laboratory (PNNL).

\vspace{15pt}
{\renewcommand{\baselinestretch}{1.3}
\footnotesize{J.~Dalmasson,  R.~DeVoe,  D.~Fudenberg,  G.~Gratta, A.~Jamil, M.~J.~Jewell, G.~Li, L.~Lupin-Jimenez, S.~Kravitz,  A.~Schubert
% ...
Drexel University, Philadelphia, PA 19104, USA.}

\footnotesize{L.~Fabris and R.~J.~Newby are with Oak Ridge National Laboratory, Oak Ridge, TN 37831, USA.}

\footnotesize{S.~Feyzbakhsh and A.~Pocar are with Amherst Center for Fundamental Interactions and Physics Department, University of Massachusetts, Amherst, MA 01003, USA.}

\footnotesize{K.~S.~Kumar, O.~Njoya and M.~Tarka are with Department of Physics and Astronomy, Stony Brook University, SUNY, Stony Brook, NY 11794, USA.}

\footnotesize{D.~S.~Leonard is with IBS Center for Underground Physics, Daejeon 34126, Korea.}

\footnotesize{J.-L.~Vuilleumier is with LHEP, Albert Einstein Center, University of Bern, Bern CH-3012, Switzerland.}

\footnotesize{I.~Badhrees's Home Institutue is King Abdulaziz City for Science and Technology, KACST, Riyadh 11442, Saudi Arabia.}

\footnotesize{G.~F.~Cao is also with University of Chinese Academy of Sciences, Beijing 100049, China.}

\footnotesize{B.~Cleveland is also with SNOLAB, Ontario, Canada.}

\footnotesize{W.~Cree is now at Canadian Department of National Defense.}

\footnotesize{S.~Delaquis is deceased.}

\footnotesize{Y.~Ito is now at JAEA, Ibaraki, Japan.}

\footnotesize{S.~Kravitz is now with Lawrence Berkeley National Lab, Berkeley, CA, USA.}

\footnotesize{A.~Schubert is now with OneBridge Solutions, Boise, ID, USA.}

\footnotesize{M.~Tarka is now with University of Massachusetts, Amherst, MA, USA.}

\footnotesize{X.~Zhang is now with Tsinghua University, Beijing, China.}}

\ifCLASSOPTIONcaptionsoff
  \newpage
\fi

\bibliographystyle{IEEEtran}
\bibliography{IEEEabrv,mybibfile}
\end{document}